\documentclass[useAMS,usegraphicx]{aa}
\usepackage{epsf,url}
\usepackage{subfigure,amssymb,amsfonts,amstext,amsgen,amsopn,amsxtra,indentfirst,lscape,times,rotating}
\usepackage{longtable}
\usepackage{pdflscape}
\usepackage{graphics}
\usepackage{keyval}
\usepackage{trig}
\usepackage[dvips]{color}
\usepackage{dcolumn}
\usepackage{bm}
\newcommand\ignore[1]{} 
\def\beq{\begin{equation}}
\def\eeq{\end{equation}}
\def\bey{\begin{eqnarray}}
\def\eey{\end{eqnarray}}

\def\kms{\, {\rm km \, s}^{-1} }
 
\def\grad{{\bf \nabla}}

\def\mnras{MNRAS}

\def\apjs{ApJ}
\def\apjl{ApJ}

\def\aap{A \& A}
\def\aj{AJ}

\def\aap{Astron. Astrophys.}

\begin{document}

\title{The dynamics of the bulge dominated galaxy NGC 7814 in MOND}
\titlerunning{NGC 7814 in MOND}

\author{G. W. Angus\fnmsep\inst{1}
\and K. J. van der Heyden\fnmsep\inst{1}
\and A. Diaferio\inst{2,3}
}
\institute{
Astrophysics, Cosmology \& Gravity Centre, University of Cape Town, Private Bag X3, Rondebosch, 7701, South Africa   
\and
Dipartimento di Fisica, Universit\`a degli Studi di Torino, Via P. Giuria 1, I-10125, Torino, Italy 
\and
Istituto Nazionale di Fisica Nucleare (INFN), Sezione di Torino, Torino, Italy
}

\date{Received date; accepted date} 
\abstract{
The bulge dominated galaxy NGC~7814 provides one of the strongest dynamical tests possible for Modified Newtonian Dynamics (MOND). Spitzer 3.6~$\mu m$ photometry fixes the bulge parameterisation and strongly constrains the properties of the subdominant stellar disk. Furthermore, the distance is known to better than 5\%, virtually eliminating it as a free parameter. The rotation curve is easily measured, since the H I (and stellar) disks are edge on, and both the receding and approaching sides agree very well.}
{To explore the agreement between the model and observed rotation curves in MOND given that the only two free parameters available are the mass-to-light ratios of the bulge and disk.} 
{We use a grid based MOND Poisson solver that accurately solves for the MOND gravity and produces our model rotation curves from a given mass distribution. The input to the Poisson solver is a 3D distribution of N particles which is generated from modelling the observed distribution of stars and gas in the galaxy.}
{By ensuring a superior fit to the radial surface brightness profile than previous works, by virtue of a double S\'{e}rsic fit to the bulge, we were able to produce excellent fits to the rotation curve with typical values for both mass-to-light ratios.}
{The model rotation curve of a mass distribution in MOND is extremely sensitive to the bulge-disk decomposition and even slight deviation from the observed mass distribution can produce large differences in the model rotation curve.}
\maketitle

\section{Introduction}
\protect\label{sec:intr}
As {H\,{\sc I~}} studies of galaxies become increasingly more deep and resolved, both spatially and in terms of their velocity, the ability to constrain Modified Newtonian Dynamics (MOND; Milgrom 1983, but for a recent review see Famaey \& McGaugh 2012) grows. Arguably, MOND surpasses theories that require galactic dark matter in explaining the dynamics of galaxies mainly due to its greater predictiveness.

The premier tool for measuring the dynamics of galaxies is the {H\,{\sc I~}} rotation curve. Once one is in possession of a galaxy's distance and luminous mass distribution, one should be able to precisely determine that galaxy's expected rotation curve and test the theory by comparing with the observed rotation curve. Unfortunately, uncertainties often get in the way of a straight-forward comparison between the measured and predicted rotation curve, whether be it in terms of the distance uncertainty, or more commonly from the so-called mass-to-light ratios of the stellar components.

The mass-to-light ratio ($M/L$) of a stellar component is the unknown relationship between the luminosity, of that ensemble of stars in a certain waveband, and its mass. If the precise luminosity function of stars for the stellar component was known, then we could exactly determine the $M/L$, but not even the initial mass function is known and multiple epochs of star formation occurring at different locations within the galaxy may cloud the result from a simple single epoch.

To combat this, state of the art photometry of galaxies is performed at infrared wavelengths such as the 2~$\mu m$-band (used by the 2MASS survey; Skrutskie et al. 2006) and 3.6~$\mu m$ (used by the SINGS survey; Kennicutt et al. 2003). These wavelengths are more sensitive to the lower mass stars, which contribute the bulk of the mass, and are less sensitive to dust extinction than are optical wavelengths. Incorporating information about the colour of the stellar component enables us to date it and make our estimation of the $M/L$.

As alluded to above, distances can also inject a large uncertainty, however, recent measurements based on locating the tip of the red giant branch in the galaxy's colour-magnitude diagram can facilitate a distance estimation with as little as five per cent error (e.g. Ferrarese et al. 2000). When one analyses the light distribution in a disk galaxy, if it is slightly inclined from being totally edge on, it is impossible to accurately determine the vertical distribution of stars and gas so one also has some freedom in fitting for the scale heights of those distributions.

In addition to the uncertainties associated with the vertical mass distribution, the $M/L$s of the various stellar components and distance, there is also the inherent uncertainty associated with how MOND transitions between the strong and weak gravity regimes. Strong and weak are relative to the new, presumably fundamental, constant of acceleration, $a_o$, that encapsulates MOND. This also has uncertainty and is taken to be around $a_o \approx 3.6 (\kms)^2pc^{-1}$. This, preferably smooth, transition can take on a wide variety of forms and can substantially alter the predicted rotation curve at intermediate values of acceleration.

Fraternali et al. (2011, hereafter FSK11) have recently presented a comprehensive analysis of a bulge dominated galaxy - NGC~7814 - with an edge on disk that has a symmetrical rotation curve (receding and approaching) which can provide an excellent test of MOND, perhaps one of the strongest tests to date. For one, the distance is measured to better than 5 per cent using the tip of the red giant branch, the galaxy has 3.6~$\mu m$ photometry and since it is edge on, it has an accurately determined vertical stellar distribution. The only tunable parameters that can make a significant difference to the fitted rotation curve are the $M/L$s of the stellar components. The reason we talk about stellar components is because NGC~7814 is a bulge dominated galaxy, but it has a stellar disk in addition to an {H\,{\sc I~}} disk.

NGC~7814 is one half of a pair of galaxies, the other being NGC~891, originally considered by van der Kruit (1983, 1995). These two are akin to the pair NGC~2403 and UGC~128 later considered by de Blok \& McGaugh (1998) in the sense that it is of fundamental importance whether galaxies with the same total mass, but different mass distribution have appropriate rotation curves in MOND.



Angus et al. (2012) presented a Poisson solver for Milgrom's QUasi-Linear formulation of MOND (QUMOND; Milgrom 2010). It takes the input from any 3D distribution of particles and accurately calculates the QUMOND gravity everywhere. This is important because strictly the typical algebraic relation of MOND used by the majority of researchers is only valid in spherical symmetry. The code was shown to be effective at fitting the rotation curves of a small sample of the THINGS survey galaxies (de Blok et al. 2008).

Here we seek to discover whether a detailed analysis of the available data on NGC~7814 and using our sophisticated tool to generate the MOND gravity from the source 3D density distribution can assist an excellent fit to the rotation curve with physically motivated parameters i.e. reasonable $M/L$s for both the bulge and the disk. In this paper the acceleration constant of MOND is always assumed to be $a_o=3.6(\kms)^2pc^{-1}$ and the interpolating function is assumed to be $\nu(y)=0.5+0.5\sqrt{1+4/y}$, where $y={|\grad \Phi_N| \over a_o}$ and $\Phi_N$ is the Newtonian potential.

\section{Bulge-Disk Decomposition}
Fits to an observed surface brightness with even mild central deviations - especially in the case of a bulge - can have a severe impact on the potential agreement between the rotation curve and MOND. Therefore, we made our own careful bulge-disk decomposition for NGC~7814. The primary piece of data we used was the radial surface brightness along the major axis. We took the data from Fig 6, right hand panel, of FSK11 and used the formula $2.5log_{10}I(R)=21.572+3.24+\mu_{3.6\mu m}$ to convert the surface brightness in magnitudes per squared arcsecond to projected luminosity $L_{\odot,3.6\mu m}pc^{-2}$, where the 3.24 corresponds to the absolute magnitude of the Sun at $3.6\mu m$.

Using a single S\'{e}rsic profile it is not possible to obtain a good match to the surface brightness profile both for $R<0.5~kpc$ and $0.5<R<1~kpc$, where the bulge is the dominant component. Additionally, the stellar disk is the dominant component of surface brightness in the disk plane for $R>1~kpc$ (see Fig~\ref{fig:sbr}), but it is also not possible to fit the surface brightness profile for $1<R<4.5~kpc$ and $R>4.5~kpc$ with a single exponential disk.

Therefore we fitted a double exponential model for the disk and double S\'{e}rsic density profile for the bulge i.e. one described by a certain S\'{e}rsic profile (or exponential disk) up to a given radius and then described by another beyond that radius. The combined, bulge and disk respectively, projected luminosity $I(R,z)=I_b(R,z)+I_d(R,z)$ was thus\\
\bey
\nonumber I(R,z=0)&=&I_{e} exp(-7.67((R/R_{e})^{1/n_{s}}-1))\\
\nonumber &&+\Sigma_{d} exp(-R/R_{d})\\
\nonumber I(R=0,z)&=&I_{e} exp(-7.67((|z|/z_{e})^{1/n_{s}}-1))\\
&&+\Sigma_{d} sech^2(|z|/z_d)
\eey
One combination of S\'{e}rsic parameters ($I_{e_i}$, $R_{e_i}$, $n_{s_i}$) is used only to fit the inner 0.5~kpc and another combination ($I_{e_o}$, $R_{e_o}$, $n_{s_o}$) is used only beyond 0.5~kpc. Also one combination of disk parameters ($\Sigma_{d_i}$, $R_{d_i}$) is used to fit the inner $4.5~kpc$ and another is used only beyond $4.5~kpc$ ($\Sigma_{d_o}$, $R_{d_o}$). We used the subscripts ``i'' and ``o'' refer to the $inner$ and $outer$ parameters, as per table~\ref{tab:par}. These parameters are not to be interpreted physically in terms of scaling relations. One can see that the former and latter combinations of parameters bracket those of FSK11, in the sense that $I_{e_i}$ is lower than the single value of the S\'{e}rsic fit performed by FSK11 and $I_{e_o}$ is larger. All fit parameters are given in table~1.


One can see in Fig~\ref{fig:sbr} that the radial surface density of the galaxy is better represented in our model (solid green line) than the single S\'{e}rsic, single exponential disk profile fit by FSK11 (dashed red line), especially below 0.3~kpc (right hand panel) and between 2 and 4.5~kpc (left hand panel). Beyond $R=4.5$~kpc, our exponential disk scale-length changes to a lower value and the surface brightness of our model goes through the centre of the data points, as does the FSK11 model. The dotted blue line of Fig~\ref{fig:sbr} shows the surface brightness of our model with only a bulge. This is simply to emphasise that the surface brightness for $R>1~kpc$ is mainly contributed by the disk, not the bulge, even though the bulge still contributes the bulk of the gravity at all radii.

In order to constrain the stellar disk scale-height and the axis-ratio of the bulge, $q=z_{e}/R_{e}$ (both of which are constant with radius), we attempted to match the contours of surface brightness taken from FSK11 Fig 6 top panel as well as the $z$-direction surface brightness profile at $R=0~kpc$. The 2D contours of surface brightness are equally well represented in both models (see Fig~\ref{fig:con}), which shows that both sets of disk parameters are acceptable. In fact, many correlated combinations of central surface density and scale-length will allow agreement, however, the scale-height cannot differ greatly from our 0.42~kpc (or FSK11's 0.44~kpc) or the outer contours will cease to agree.

So that we can solve for the rotation curve of our model galaxy, we must produce a 3D model of the galaxy. In order to minimise the complexity, we only allow for the z-direction variation in terms of the axial ratio, $q$, rather than a fitting for a whole new set of S\'{e}rsic parameters. Nevertheless, the data are relatively well adhered to (see Fig~\ref{fig:sbz}). 
\subsection{Total luminosity}
Given this special case of a completely edge on disk, the luminosity must be inferred by integrating the projected luminosity
\bey
\nonumber L_d&=&[\int_0^{4.5~kpc}  2\Sigma_{d_i}exp(-\hat{R}/R_{d_i}) d\hat{R} +\\
\nonumber&&\int_{4.5~kpc}^{\infty}  2\Sigma_{d_o}exp(-\hat{R}/R_{d_o}) d\hat{R}] \times\\
\nonumber&& [ \int_{-\infty}^{\infty} sech^2(\hat{z}/z_d)d\hat{z} ]\\
&=&8\times10^9 L_{\odot,3.6\mu m}.
\eey
The bulge luminosity must be integrated numerically and is found to be $L_b=10^{11} L_{\odot,3.6\mu m}$, which is 43\% higher than FSK11 bulge luminosity.

It is worth remarking that these parameters, both in the text above and table~\ref{tab:par} are all projected parameters. $R$ is the major axis which is in the horizontal plane of the sky and $z$ is the minor axis and is in the vertical plane of the sky.

\subsection{Distance and Rotation Curve}

The distance to NGC~7814 used for presenting the analysis in FSK11 was 14.6~Mpc, which is well within the errors of the distance derived by Radburn-Smith et al. (2011), as part of the GHOSTS survey, which is $14.4^{+0.7}_{-0.6}~Mpc$ (cf. their table~5). For ease of comparison, the distance is always assumed to be 14.6~Mpc in this paper, it is not a free parameter.

The rotation curve we use for comparison with the models is the mean of the approaching and receding rotation curves in the position velocity diagram presented in Fig~3 of FSK11 (private communication with the authors of FSK11).


\section{Our model}
\protect\label{sec:ourm}
We generated a $512^3$ particle realisation of NGC~7814 where we dedicated half of the particles to the bulge and a quarter each to the {H\,{\sc I~}} and stellar disks. The luminosity profile was strictly adhered to using the surface brightness models advocated above, as was the {H\,{\sc I~}} surface density, found in FSK11 Fig~7 right hand panel, by making an Abel transformation from our analytical 2D models to numerical 3D models, which we stored. We assume the galaxy is symmetrical around the minor axis.

For distributing the particles of the bulge, we used a standard rejection technique: we drew random 3D radii from an exponential distribution and simply interpolated through the data points of our numerical 3D model (created by the Abel inversion) to find the 3D density at that given radius. Then we decided to keep the particle if the next random number was less than the ratio of the value of the density at that radius of the numerical density and the analytical one. We then randomly assigned $x$, $y$ and $z$ spherical coordinates but multiplied the spherical $z$ coordinate by the axis ratio $q$ to create the flattening. For the disks we distributed the particles in $z$ according to a $sech^2(z)$ profile: the gas with a 0.1~kpc scale-height and stars using the scale-height given in table~\ref{tab:par}. The $x$ and $y$ coordinates were chosen as for the bulge, except our analytical distribution was 2D; meaning the cumulative distribution function differs from the 3D case.

Using the QUMOND Poisson solver code presented in Angus et al. (2012), we solved for the gravity in the midplane of the disk and used this to produce the model values of the rotation speed as a function of projected radius, $V_{model}(R)$.
 
Our task then was to explore the parameter space of the free parameters in order to elucidate the agreement of MOND with the observed rotation curve. The only parameters free to be varied were the mass-to-light ratios of the bulge and disk. We parameterise the goodness of fit as $\chi^2=\Sigma_{i=1}^n\left(  {V_{model}(R_i) - V_{obs}(R_i) \over \sigma_{obs}(R_i)}\right)^2$, where $n=18$.

\section{Results}
In Fig~\ref{fig:contour} we show contours of goodness of fit, $\chi^2/n$ for the 2D parameter space defined by the mass-to-light ratios of the bulge and disk ($x$ and $y$ axes respectively). We find that the best fits ($\chi^2/n < 0.35$) occur for values from $M/L_D=0.6$ to $1.1$ and that values as low as 0.3 (and high as 1.4) can still produce excellent fits with $\chi^2/n < 0.45$. To achieve $\chi^2/n < 0.45$ the mass-to-light ratio of the bulge, $M/L_B$, is required to be in a significantly tighter range $0.58$ to $0.685$.


In Fig~\ref{fig:rc} we plot the rotation curves of four models against the data points measured by FSK11. The black lines use our parameterisation of the bulge and disk and use mass-to-light ratios of (0.66,0.7) and (0.685,0.3) for the solid and dashed lines respectively.
\subsection{Comparison with other work}
\begin{figure*}
\centering
\subfigure{
\includegraphics[angle=0,width=8.50cm]{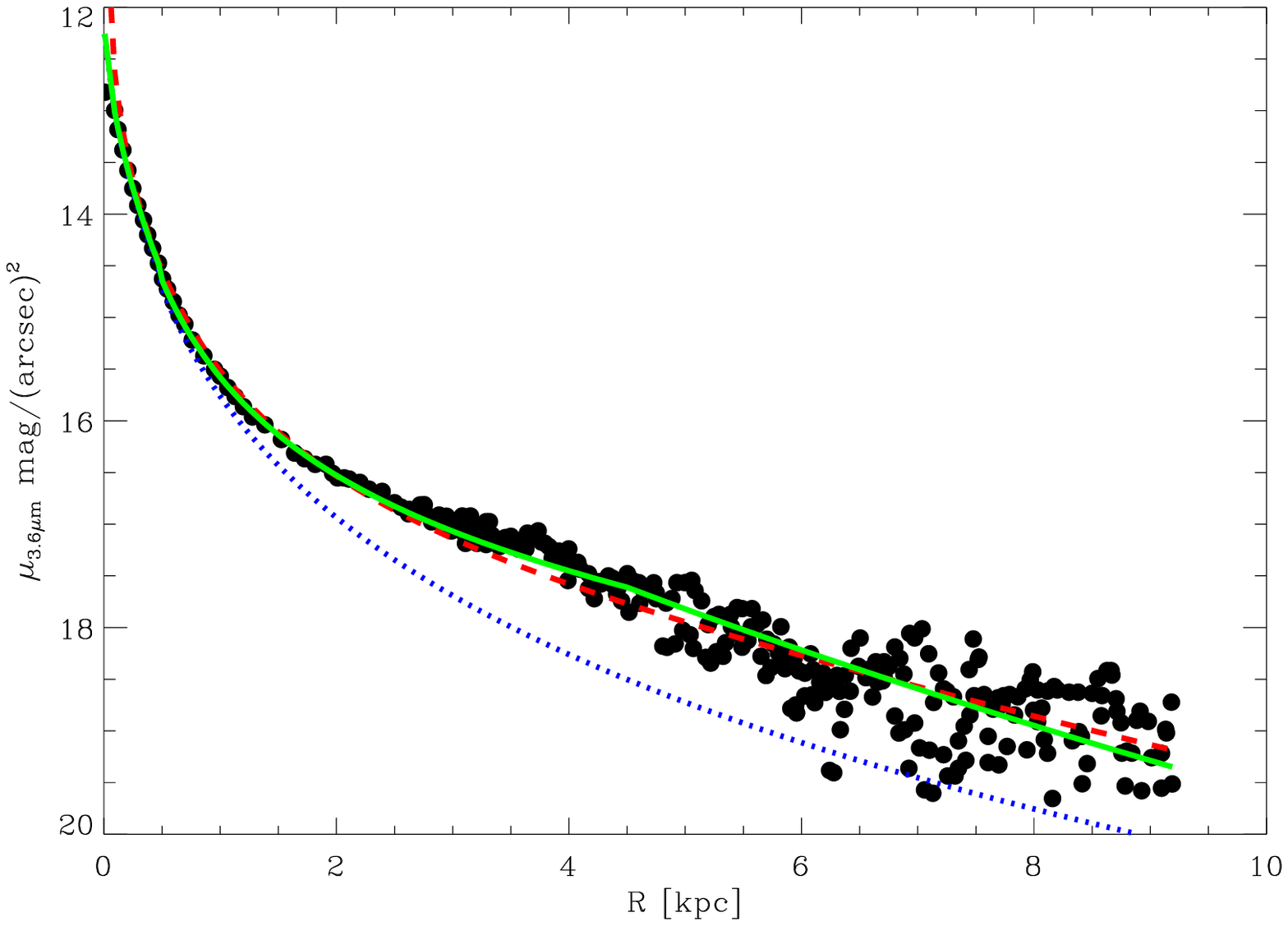}
}
\subfigure{
\includegraphics[angle=0,width=8.50cm]{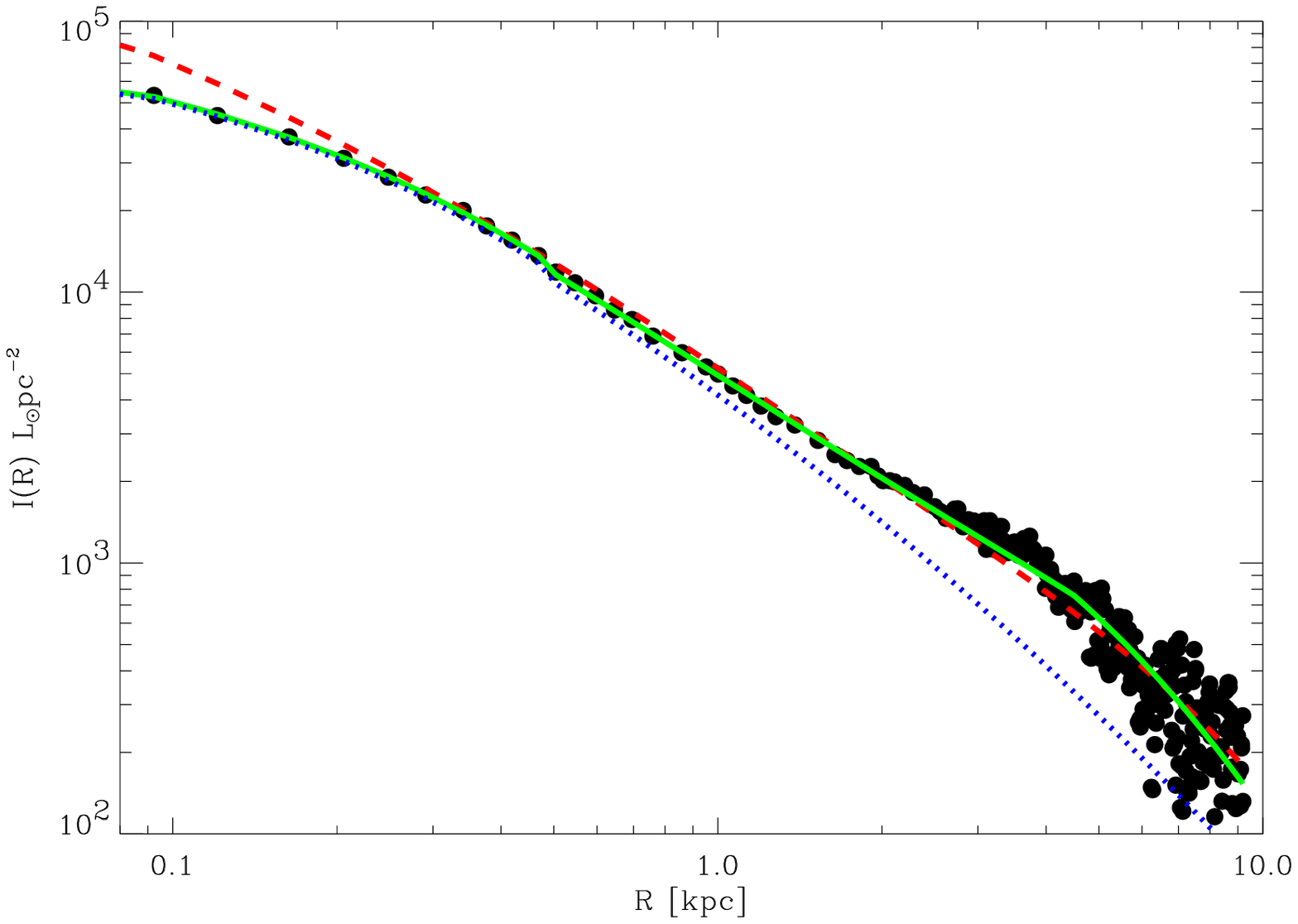}
}\\
\caption{In the left hand panel we plot surface brightness against projected radius along the major axis of NGC~7814. The data points come from FSK11 and their fitted model is the red dashed line. Our model, using the combined luminosity of the bulge and disk, is the green solid line and the blue dotted line is the surface brightness of our fitted bulge only. The right hand panel shows the projected luminosity against projected radius along the major axis. The fitted models are the same as in the left hand panel.}
\label{fig:sbr}
\end{figure*}

\begin{figure}
\includegraphics[angle=0,width=8.50cm]{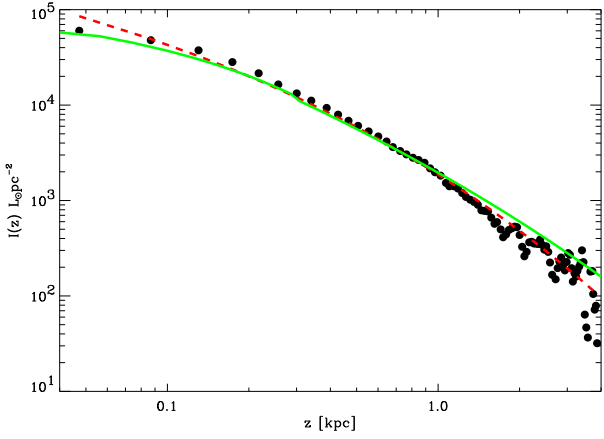}
\caption{Here we plot the projected luminosity against projected radius along the minor axis of NGC~7814. The data points come from FSK11 and their fitted model is the red dashed line. The green solid line is our model, which is not a fit, but rather is the projected luminosity profile fit of the major axis squashed by the axis ratio, $q$, which is determined from fitting the surface brightness contours of Fig~\ref{fig:con}.}
\label{fig:sbz}
\end{figure}

\begin{figure}
\includegraphics[angle=0,width=8.50cm]{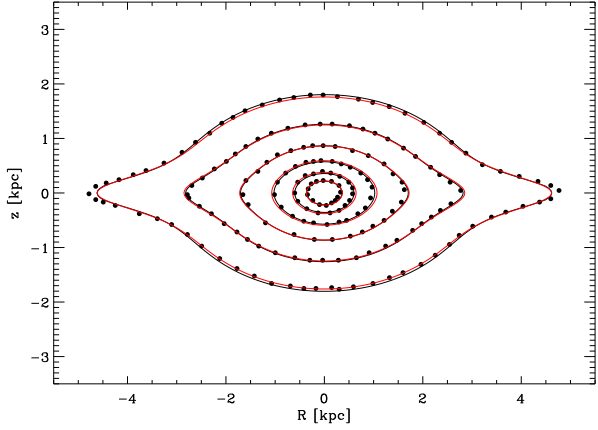}
\caption{Here we show the fits to the surface brightness contours of NGC~7814 presented in FSK11. The models are the closed curves and the filled circles are the observed contours. The black lines correspond to our model and the red lines to that of FSK11. The parameters of the two models are given in table~\ref{tab:par}.}
\label{fig:con}
\end{figure}

FSK11 performed a bulge-disk decomposition which separated the stellar distribution into a bulge and thin disk. To test MOND, they used the velocity profiles derived for each component (i.e. they solved Poisson's equation to convert surface brightness to gravity and hence circular speed) and they found that agreement between the model rotation curve and the measured one could only be found if $M/L_D=4.6$; an unfeasibly large disk mass-to-light ratio. In Fig~\ref{fig:sbr} the red dashed line corresponds to the FSK11 fit to the surface brightness, which is not as good a fit as our model to the inner 4.5~kpc. This is simply because NGC~7814 has a relatively complicated surface brightness profile and a double S\'{e}rsic bulge and double exponential disk is required to trace the profile over the full range.

In order to compare the surface brightness models of FSK11 we need to know their parameterisations. The difference between projected and deprojected parameters is never clearly distinguished in FSK11, but we assume the ``$r_e$'' they use is the projected S\'{e}rsic radius. On the other hand, the surface density they give for the disk is almost certainly projected in $z$ and deprojected along the line of sight (i.e. as seen from above the disk - face on, whereas our true perspective is that of an edge on disk), even though the disk scale-length given is presumably the projected one. Therefore, using the luminosity of their disk, we re-derived the parameters for their model and have added them to table~\ref{tab:par}.

\begin{figure}
\includegraphics[angle=0,width=8.50cm]{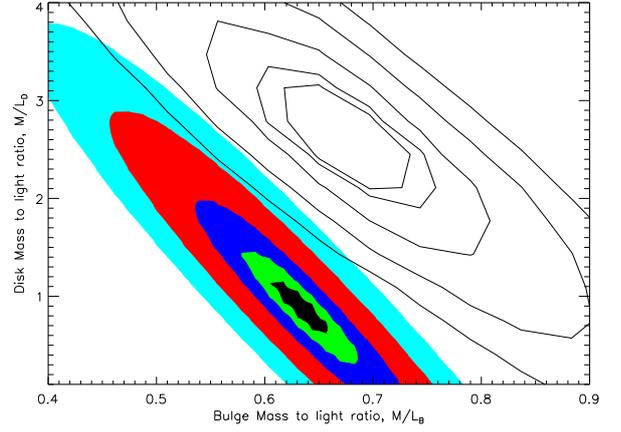}
\caption{Here we show the contours of goodness of fit, $\chi^2/n$ for the 2D parameter space defined by the mass-to-light ratios of the bulge and disk ($x$ and $y$ axes respectively) found by fitting the rotation curve of NGC~7814. The filled contours are found using our parameterisation of the bulge and disk and the plain contours are using FSK11's parameterisation. Both sets of contours use the same levels and both use the simple $\nu$ function of QUMOND. The central contour has a value $\chi^2/n=0.35$ and the subsequent contours are 0.45, 0.8, 2 and 4.}
\label{fig:contour}
\end{figure}

Using these parameters we were able to follow the same procedure as describe in the beginning of \S\ref{sec:ourm} to generate a 3D model of NGC~7814. We then fully explored the parameter space defined by the mass-to-light ratios of the two stellar components. The plain contours in Fig~\ref{fig:contour} are using FSK11's parameterisation, for which the best fit clearly requires a large $M/L_D$ of around 2.65 and it must be larger than 1.5 in order to have a $\chi^2/n < 0.8$. This is incompatible with the expected mass-to-light ratios of star forming regions in the 3.6~$\mu m$ band, which should have values lower than that of the bulge, which itself should have a value around 0.7. The best fit mass-to-light ratio for their bulge is 0.61-0.73, which is not too dissimilar to our own.

\begin{figure}
\includegraphics[angle=0,width=8.50cm]{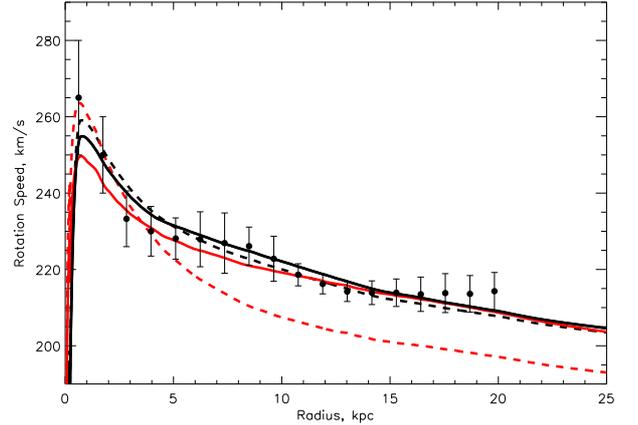}
\caption{Here we show the rotation curves of four fits to the measured rotation curve of NGC~7814 (data points from FSK11). The black lines use our parameterisation of the galaxy and the red lines use that of FSK11. The bulge and disk mass-to-light ratios are (0.66,0.7), (0.685,0.3), (0.66,2.7) and (0.75,0.7) for the linetypes black solid, black dashed, red solid and red dashed respectively.}
\label{fig:rc}
\end{figure}

In Fig~\ref{fig:rc} we have plotted the best fit rotation curve using the FSK11 model (red solid line), with $M/L_B=0.66$ and $M/L_D=2.65$, against the measured rotation curve as well as a fit using a more regular value for the disk mass-to-light ratio, $M/L_B=0.75$ and $M/L_D=0.7$ (red dashed line). The best fit is, as expected, as good a fit as the solid black line (our best fit model), but the red dashed line shows that using a regular $M/L_D$ leads to a very poor fit.

We emphasise that it is the difference between our brightness profile for the bulge and that of FSK11 that allows us to reduce the mass-to-light ratio of the subdominant stellar disk, not the different parameterisation of our disk and theirs. 

\begin{table*}
\begin{tabular}{|c|c|c|c|cc|c|c|c} \hline
\small range &  \small $\sqrt{R^2+z^2/q^2} <0.5~kpc$   & \small $\sqrt{R^2+z^2/q^2} >0.5~kpc$   &-------  & \small  & $R<4.5~kpc$ & $R>4.5~kpc$& ------------------------------------ \\ \hline
\end{tabular}
\begin{tabular}{|c|ccc|ccc|c|cc|cc|c|cc|} \hline
\small model & \small $I_{e_i}$ & \small $R_{e_i}$ & \small $n_{s_i}$ & \small $I_{e_o}$ &  \small $R_{e_o}$   & \small $n_{s_o}$ & \small $q$  & \small $\Sigma_{d_i}$ & $R_{d_i}$ & \small $\Sigma_{d_o}$ & $R_{d_o}$ &$z_d$&$L_b$&$L_d$\\ \hline
	     &$L_{\odot}pc^{-2}$&$kpc$    &		      &$L_{\odot}pc^{-2}$&$kpc$       &		  &	      &$L_{\odot}pc^{-2}$&$kpc$&$L_{\odot}pc^{-2}$&$kpc$&$kpc$&$10^{10}L_{\odot}$&$10^{10}L_{\odot}$\\ \hline
{\bf A12} & 60.05 & 3.82  & 1.747  & 9000 & 0.578 & 5.747 & 0.61  & 900 & 6.0& 2250 & 2.7 & 0.42 & 10.0 & 0.8 \\ \hline
{\bf FSK11} & 1120 & 2.16  & 4  & 1120 & 2.16 & 4 & 0.61 & 1200 & 4.26& 1200 & 4.26 & 0.44 & 7.0 &0.84 \\ \hline
\end{tabular}
\caption{Here we list the parameters of the bulge and disk in our fitted model (referred to as A12) and the FSK11 fit. Note the ranges of applicability for the A12 model. The subscripts ``i`` and ''o`` refer to the $inner$ and $outer$ fitted parameters, both for the bulge and disk. All luminosities are in the 3.6~$\mu m$ band.}
\protect\label{tab:par}
\end{table*}

\section{Conclusion}
The bulge dominated galaxy NGC~7814 has an edge on stellar disk and a regular H~I disk. The bulge-disk decomposition performed by FSK11 suggested that an impossibly large mass-to-light ratio for the stellar disk was required in order to fit the rotation curve of the galaxy. Here we used two S\'{e}rsic profiles to model the projected luminosity of the bulge such that one profile is used below 0.5~kpc and the other is used beyond 0.5~kpc. This allows a superior fit to the observed surface brightness, especially below 0.3~kpc and between 2 and 4.5~kpc. We then deprojected the projected luminosity models for bulge and disk, modelled the galaxy with a 3D particle distribution and used the code of Angus et al. (2012) to compare the observed and modelled rotation curves across the 2D parameter space comprising the mass-to-light ratios of the bulge and disk. We made the same analysis using the models of FSK11.

It was found that with the FSK11 model, excellent fits to the rotation curve could only be found using very large mass-to-light ratios for the disk (larger than 1.5 in order to have a goodness of fit $\chi^2/n < 0.8$). With our model it was found that $\chi^2/n < 0.45$ was achievable for disk mass-to-light ratios in the range 0.3-1.4 and that the goodness of fit was most strongly constrained by the mass-to-light ratio of the bulge - which should be between 0.58-0.685 for the same level of goodness of fit. This mass-to-light ratio for the ancient population of stars in the bulge is precisely in the range expected by stellar population models (see Bell \& de Jong 2001).

Here we have re-analysed the required properties of one of the most pristine test galaxies of Modified Newtonian Dynamics. There is virtually no room for manoeuvring with respect to the galaxy's inclination, distance, structure and we keep all parameters associated with MOND at the standard values used in the literature. We found that by forcing a more stringent adherence to the luminosity profile of the bulge than previously used, it is possible to, firstly, make an excellent fit to the rotation curve of the galaxy and, in doing so, isolate the two free parameters required by the fit. These two free parameters, namely the  mass-to-light ratios of the bulge and disk, are perfectly compatible with the predicted vales.

Our results show that a proper comparison between observed rotation curves and the MOND expectations requires extreme care when performing bulge-disk decompositions, especially for the central regions.

\section{Acknowledgments} The authors acknowledge the time and effort invested by Fraternali, Sancisi \& Kamphuis (2011) in making all observations that we make use of in the paper. They also thank the referee Stacy McGaugh for his wise suggestions that improved the readability and robustness of the paper. GWA acknowledges the ``Modified Gravities Approaches to the Dark Sector'' workshop organised by Benoit Famaey in Strasbourg 2010, where the issue of NGC~7814 was first discussed. GWA is supported by the Claude Leon Foundation and a University Research Committee Fellowship from the University of Cape Town. Both GWA and KvdH are funded by the National Research Foundation of South Africa and in particular the NRF special award for Y-rated researchers is gratefully acknowledged. AD acknowledges partial support from the INFN grant PD51 and PRIN-MIUR-2008 grant \verb"2008NR3EBK_003" ``Matter-antimatter asymmetry, dark matter and dark energy in the LHC era''.


\begin{thebibliography}{}

\bibitem[]{}
 Angus, G. W., van der Heyden K. J., Famaey, B., Gentile, G., McGaugh, S. S. \& de Blok, W. J. G. 2012, MNRAS, 421, 2598

\bibitem[]{}
Bell, E. F., de Jong, R., 2001, ApJ, 550, 212

\bibitem[]{}
de Blok, W.J.K., McGaugh, S.S, 1998, ApJ, 508, 132

\bibitem[{{de Blok} {et~al.}(2008){de Blok}, {Walter}, {Brinks},
  {Trachternach}, {Oh}, \& {Kennicutt}}]{deblok08}
{de Blok}, W.~J.~G., {Walter}, F., {Brinks}, E., {Trachternach}, C., {Oh}, S.,
  {Kennicutt}, R.~C., 2008, \aj, 136, 2648

\bibitem[]{}
Fraternali, F., Sancisi, R. \& Kamphuis, P. 2011, A \& A, 531, A64 (FSK11)

\bibitem[]{}
Radburn-Smith, D., et al., 2011, ApJS, 195, 18

\bibitem[{{Famaey} \& {McGaugh}(2011)}]{famaey12}
Famaey, B., McGaugh, S., 2012, Living Reviews in Relativity, 15, 10

\bibitem[]{}
Ferrarese, L., et al., 2000, APJS, 128, 431

\bibitem[]{}
Kennicutt, R. C., et al., 2003, PASP, 115, 928

\bibitem[{{Milgrom}(2010)}]{milgrom10}
Milgrom, M., 2010, \mnras, 403, 886

\bibitem[]{}
Skrutskie, M.~F., et al., 2006, AJ, 131, 1163
\bibitem[]{}
van der Kruit, P. 1983, PASAu, 5, 136
\bibitem[]{}
van der Kruit, P. C. 1995, in Stellar Populations, ed. P. C. van der Kruit, \& G.
Gilmore (Dordrecht: Kluwer Acad. Publ.), IAU Symp., 164, 205

\end{thebibliography}
\end{document}